\documentclass[11pt]{article}
\usepackage{epsf}
\usepackage{graphics}
\usepackage{epsfig}
\usepackage{rotating}
\catcode`\@=11

\@addtoreset{equation}{section}
\def\theequation{\arabic{section}.\arabic{equation}}
\catcode`\@=11

\def\section{\@startsection{section}{1}{\z@}{3.5ex plus 1ex minus
   .2ex}{2.3ex plus .2ex}{\large\bf}}

\def\eqnarray{\let\@currentlabel=\theequation\refstepcounter{equation}
    \global\@eqnswtrue
    \global\@eqcnt\z@\tabskip\@centering\let\\=\@eqncr
    $$\halign to \displaywidth\bgroup\@eqnsel\hskip\@centering
      $\displaystyle\tabskip\z@{##}$&\global\@eqcnt\@ne
       \hfil${{}##{}}$\hfil
      &\global\@eqcnt\tw@ $\displaystyle\tabskip\z@{##}$\hfil
       \tabskip\@centering&\llap{##}\tabskip\z@\cr}
\def\lefteqn#1{\hbox to 4\arraycolsep{$\displaystyle #1$\hss}}
\def\thesection{\arabic{section}.}

\def\appendix{\setcounter{section}{0}
        \def\thesection{Appendix.}
        \def\theequation{\Alph{section}.\arabic{equation}}}

\long\def\@makefntext#1{\parindent 0cm\noindent
\hbox to 1em{\hss$^{\@thefnmark}$}#1}
\def\IR{{\hbox{{\rm I}\kern-.2em\hbox{\rm R}}}}
\def\IH{{\hbox{{\rm I}\kern-.2em\hbox{\rm H}}}}
\def\IC{{\ \hbox{{\rm I}\kern-.6em\hbox{\bf C}}}}
\def\IZ{{\hbox{{\rm Z}\kern-.4em\hbox{\rm Z}}}}

\newcommand{\beq}{\begin{equation}}
\newcommand{\be}{\begin{equation}}
\newcommand{\eeq}{\end{equation}}
\newcommand{\ee}{\end{equation}}
\newcommand{\bea}{\begin{eqnarray}}
\newcommand{\eea}{\end{eqnarray}}
\newcommand{\bean}{\begin{eqnarray*}}
\newcommand{\eean}{\end{eqnarray*}}
\newcommand{\ba}{\beq\begin{array}{lll} }
\newcommand{\ea}{\end{array}\eeq}

\def\dag{\dagger}

\def\IC{ {\rm l\hspace{-1.2ex}C} }    %  mine is better....
\def\IZ{{\hbox{{\rm Z}\kern-.4em\hbox{\rm Z}}}}
\def\IR{{\hbox{{\rm I}\kern-.2em\hbox{\rm R}}}}
\begin{document}                           %
                                                                 %
    
%%%%%%%%%%%%%%%%%%%%%%%%%%%%%%%%%%%%%%%%%%%%%%%%%%%%%%%%%%%%%%%%%%%%%%%

%%%%%%%%%%%%%%%%%%%%%%%%%%%%%%%%%%%%%%%%%%%%%%%%%%%%%%%%%%%%%%%%%%%%%%%%%%
%%%%%%%%%%%%%%%%%%%%%%%%%%%%%%%%%%%%%%%%%%%%%%%%%%%%%%%%%%%%%%%%%%%%%%%%%

\begin{titlepage}
\vspace{.5in}
%\begin{flushright}
%Los Alamos  \\
%IEEC-Jun-97\\
%gr-qc/???????\\
%June 1997\\ {\tiny (This version: \today) }
%\end{flushright}
\vspace{.5in}
\begin{center}

{\huge\bf Spherical Harmonics Interpolation,  \\[.2in]  Computation of Laplacians\\[.2in]   and Gauge Theory  }\\\vspace{1.2in}

\includegraphics[width=7cm]{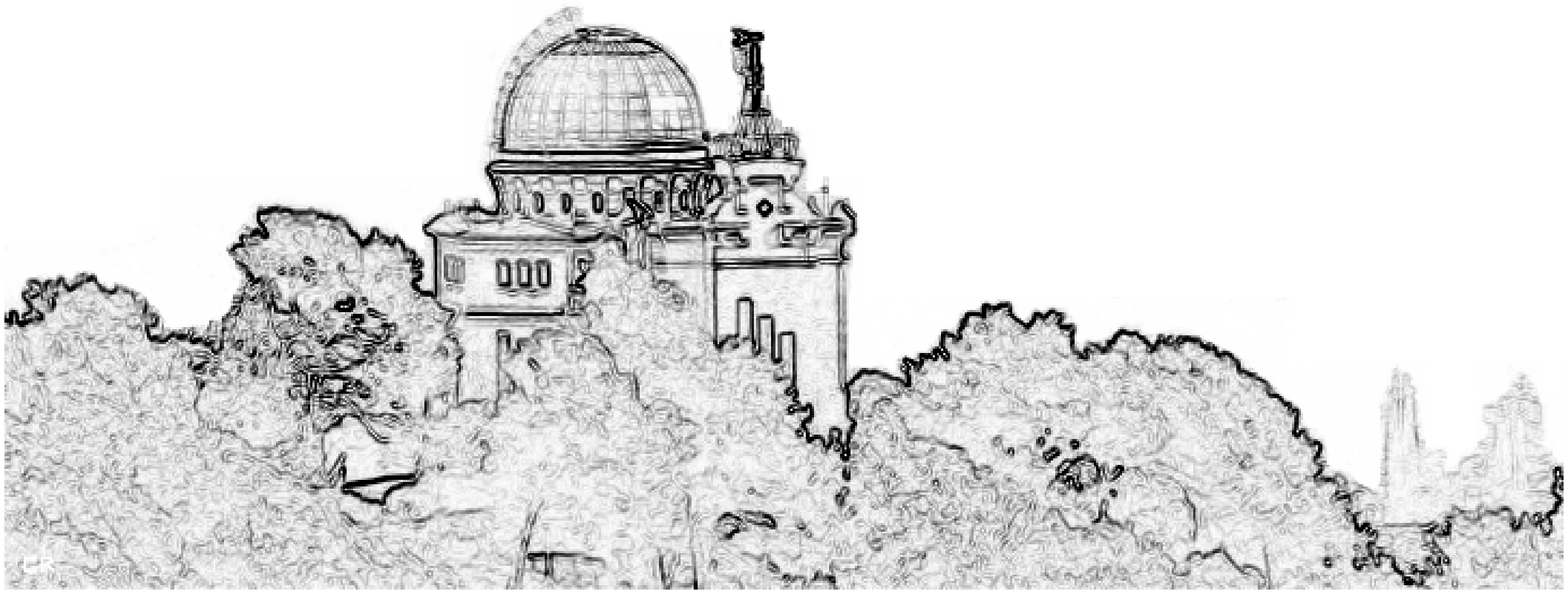} \\ \vspace{.15in}

\includegraphics[width=4cm]{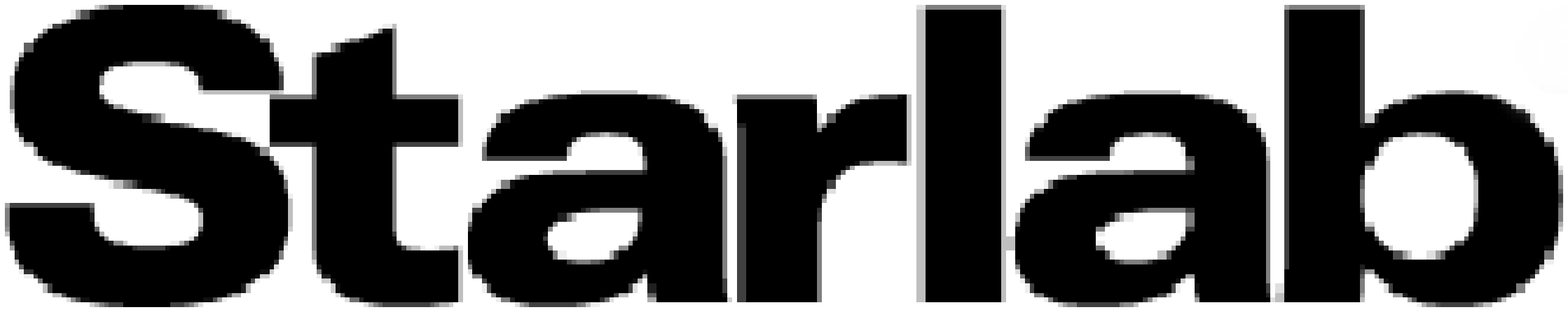}\\
\vspace{1cm}
{\large {\bf Starlab Research  Knowledge Nugget 2001-10-25} \\ Created November 1, 2001. Revised May 2002.} \\
{\large \bf Status: Public}\\
%{\large  Distribution: Starlab collaborators: Carles Grau, Josep Marco}\\
\vspace{1in} 
%{Emil ~M{ottola}\footnote{\it email: emil@pion.lanl.gov}\\
%        {\small\it Theoretical Division Group, T-8, MS-B285 }\\
%       {\small\it Los Alamos National Laboratory}\\
%        {\small\it Los Alamos, NM 87545, USA}}\\
% \vspace{1ex}
   {Giulio ~R{uffini\footnote{\it email: giulio@starlab-bcn.com}, Josep Marco, Carles Grau}\\
        {\small\it Starlab Barcelona S.L., \today}\\
        {\small\it Edifici de l'Observatori Fabra, C. de l'Observatori s.n.}\\
%{\small\it Edif. Nexus, 204} \\
{ \small\it Muntanya del Tibidabo, 08035 Barcelona, Spain \\ }
{ \small\it Tel://+34 93 254 03 66, http://starlab.es \\ }
        } 
 \end{center}

 \vfill

\clearpage
%%%%%%%%%%%%%%%%%%%%%%%%%%%%%%%%%%%%%%%%%%%%%%%%%
\begin{center}
{\large\bf Abstract}
\end{center}
\begin{center}
\begin{minipage}{5.4in}  The aim in this note is to define an algorithm to carry out minimal curvature spherical harmonics interpolation, which is then used to calculate the Laplacian for multi-electrode EEG data analysis.  The approach taken is to respect the data. That is, we implement a minimal curvature condition for the interpolating surface  subject to the constraints determined from the multi-electrode data. We implement this approach using spherical harmonics interpolation. In this elegant example we show that minimization requirement and constraints complement each other to fix all degrees of freedom automatically, as occurs in gauge theories. That is, the constraints are respected, while only the orthogonal subspace minimization constraints are enforced.  
As an example, we discuss the application to interpolate control data and calculate the temporal sequence of laplacians from an EEG  Mismatch Negativity (MMN) experiment (using an implementation of the algorithm in IDL).   \end{minipage}
    \end{center} 
\end{titlepage}

\tableofcontents

\clearpage

%%%%%%%%%%%%%%%%%%%%%%55
%%%%%%%%%%%%%%%%%%%%%%%%%%%%%55

\section{Minimization of Curvature with constraints}
This paper focuses on an algorithm for minimal curvature interpolation on the surface of a sphere (that is, a 2D problem). The interpolation will then be used for a Laplacian calculation. This is straightforward if spherical harmonics are used for the interpolation. \medskip

There are many ways to interpolate data on a surface. Here we choose to respect the original data points and minimize the curvature of the interpolating surface. Note that this may not always be a good idea, as the data may itself be ``contradictory'' due to noise impact. Nonetheless, in the present situation this is not an issue. We will be dealing with scalp voltage measurements obtained from an even grid of electrodes. No matter how noisy the data is, there always exists an interpolation solution for a given set of measurements (e.g., there will not be two contradictory measurements at the same electrode site). \medskip

 We will thus impose full respect for the original data to fix part of the interpolation parameter space, and fill the rest using the constraint of minimal curvature.
The goal is to carry out an interpolation with
\begin{equation}
v(\psi, \theta )=\sum_{l=0}^{N}\sum_{m=-l}^{l}a_{lm}\, Y_{l,m}(\theta , \psi )
\end{equation}
minimizing the total curvature:
\begin{equation}
\int dS \, |\Delta \Phi|^2= \Phi ([a_{lm}])=\sum_{l=1}^{N}\sum_{m=-l}^{l}l^2(l+1)^2|a_{lm}|^2.
\end{equation}
The Laplacian, in spherical harmonics, is given by the simple relation
\begin{equation}
\Delta v(\psi, \theta )=-{1 \over r^2}\sum_{l=0}^{N}\sum_{m=-l}^{l} l(l+1)\, a_{lm}\, Y_{l,m}(\theta , \psi ),
\end{equation}
where $r$ is the radius of the spherical head (we used 11.36 cm for an average head).  \medskip

Since the array $v$ is real, and since $(Y_{l,m})*= Y_{l,-m}$, it follows that $(a_{l,m})*= a_{l,-m}$. This point is important if we want to write the problem using complex variables and treating $a_{l,m}$ and $(a_{l,m})^* $ as independent quantities. We have to write the problem minimizing something that carries this symmetry forth. Thus, 
\beq
\chi^2= a^\dag\cdot  B a + (v-Ma)^\dag \cdot \lambda+\lambda^\dag \cdot(v-Ma),
\eeq
where $\dag$ denotes hermitean conjugation, is a valid minimizing functional. The fact that the interchange of $a$ and $a^*$ leaves $\chi$ invariant means that both will be treated equally and extremization with respect to either will yield the same equation, albeit in complex conjugate form. \medskip

The code implementation of this is a bit more complicated for and will forego for now the complex approach. In \cite{recipes} there is an interesting section explaining how to translate a complex problem into a real one. \medskip

To translate the problem to the real domain, here we work with the expansion
\begin{equation}
v(\psi, \theta )=\sum_{l=0}^{N}  \left[ a^r_{l0}\, Y_{l,0}(\theta , \psi ))+ \sum_{m=1}^{l}a^r_{lm}\, \mbox{Re}[Y_{l,m}(\theta , \psi )] + a^i_{lm}\, \mbox{Im}[Y_{l,m}(\theta , \psi )] \right]
\end{equation}
Here $v$ is the array of voltage values at each electrode position. In the present work we will have 31 such electrode values.
The equation we want to solve (the hard constraints) can be written in the form
\begin{equation}
\label{eqmain}
v=M·a.
\end{equation}
To give a pictorial description of this matrix, M has rows of the form $[Y_{l,m}(\theta_i , \psi_i )]$:
$$ M=
\left(\begin{array}{ccccc}
Y_{0,0}(\theta_1 , \psi_1 )& Y_{1,0}(\theta_1 , \psi_1 )) & \mbox{Re}[Y_{1,1}(\theta_1 , \psi_1 )] & \mbox{Im}[(Y_{1,1}(\theta_1 , \psi_1 )]& .....\\
Y_{0,0}(\theta_2 , \psi_2 )& Y_{1,0}(\theta_2 , \psi_2 )) & \mbox{Re}[Y_{1,1}(\theta_2 , \psi_2 )] & \mbox{Im}[(Y_{1,1}(\theta_2 , \psi_2 )]& .....\\
...&...&...&...&.....\\
...&...&...&...&.....\\
Y_{0,0}(\theta_{31} , \psi_{31} )& Y_{1,0}(\theta_{31} , \psi_{31} ) & \mbox{Re}[Y_{1,1}(\theta_{31} , \psi_{31} )] & \mbox{Im}[(Y_{1,1}(\theta_{31} , \psi_{31} )]& .....
\end{array}
\right)
$$

We will use lower case letters to denote vectors (such as $a$ and $\lambda$, our unknowns, and $v$, the potential data vector, a 31-vector). Then we have to be extra careful with the dimensions of our matrices. 

Let us refer to the original equation $
M\, a=v.
$
These are in fact 31 equations for an infinite number of unknowns---not a promising perspective.  So we change strategy! The goal in our game is to minimize the curvature
\beq 
C ={1\over 2} a^t \cdot B a ,
\eeq
where $B$ is a symmetric matrix, while respecting the data (the constraints).  Note that without the constraints in the previous equation, the solution to this minimization problem is simple: $a_0$ is free, the rest must be all zero. We now that $a_0$ represents the mean voltage, so it is well determined by the constraints. The problem we are posing is therefore well defined: there exists a unique solution.
\subsection{The functional and the equations}
Our problem now is to minimize
\beq \label{eqminimize}
\chi ={1\over 2} a^T\cdot  B a + \lambda^T \cdot(v-Ma)
\eeq
with respect to the unknowns $a$ and $\lambda$.

%\section{The players and the equations}
These are the players, and we need to present them. First of all, $a$ is the spherical harmonics coefficient tensor, the object we are really after. This, in principle, is a $\infty$-array.  Then there is $v$, the aforementioned 31-vector of voltage data values measured at a fixed time t.  Now, this means $M$ is a 31x$\infty$ matrix (the notation is "output"x"input") dimensions. Now, let us summarize:
\bean
a                       & \in &\IR ^ \infty \\
v, \lambda              &\in& \IR^{31} \\
M: \IR ^ \infty         &\longrightarrow& \IR^{31}\\
M^T M : \IR ^ \infty    &\longrightarrow& \IR^{\infty} \\
 M M^T : \IR ^{31}      &\longrightarrow& \IR^{31} \\
B: \IR^\infty           &\longrightarrow & \IR^{\infty}
 \eean
To proceed, let us expand Equation \ref{eqminimize}:
\bea
\chi &=&{1\over 2} a^T\cdot  B a + \lambda^T \cdot(v-MA) \\
     &=& {1\over 2} a_{i} B_{ij} a_j + \lambda_i (v_i -M_{ij} a_j ),
\eea
and now differentiate to get the stationary point:
\bea 
{\partial \chi \over  \partial a_k} &=& {1\over 2} \delta_{ik} B_{ij} a_j + {1\over 2} a_i B_{ik} - \lambda_i M_{ij} \\
                                    &=&  a_i B_{ik} - \lambda_i M_{ik}.
\eea
Differentiating with respect to $\lambda$ yields again Equation \ref{eqmain}. Our new equations are now
\bea \label{newequations}
Ba&=&M^T\lambda \\
Ma &=& v.
\eea
Now we have to solve them. The main point is to be careful with the dimensions of all the objects---the situation is a bit trickier than usual.

\subsection{Solving the equations}
Let us now add $M^T$ times the second equation to the first one (note that, perhaps for the first time, we are not adding pears to apples):
\beq
(B+M^TM)a=M^T(\lambda +  v).
\eeq
This step is important, now we have on the left a matrix of full rank.

Now, from the second Equation in \ref{newequations} we see (assuming that $MM^T$ has full rank, as it should and we shall see below)
\beq
\lambda = (MM^T)^{-1} MBa.
\eeq
This we can take and plug into the previous result,
\beq
(B+M^TM)a=M^T (MM^T)^{-1} M Ba + M^T v.
\eeq
Hence
\beq \label{eq:main}
\left[B+M^TM -M^T (MM^T)^{-1} MB  \right] a=M^T b, 
\eeq
an important result, 
and
\beq
 a=\left[B+M^TM -M^T (MM^T)^{-1} MB  \right]^{-1}M^T b \equiv Q_{n\times n}^{-1}M^T b, . 
\eeq
with 
\beq 
Q_{n\times n}=\left[B+M^TM -M^T (MM^T)^{-1} MB  \right].
\eeq
Now, can be this be simplified? Conceptually yes. The operator $ M^T (MM^T)^{-1} M $ is a projection operator. Let us describe it in more detail because this situation is a nice example between constraints and minimization requirements.

\subsection{$M$ and SVD}
To try to see how to simplify, we will use the Singular Value  decomposition. Recall that a  very powerful conceptual and practical  tool  for studying  this problem  is provided by the Singular Value
 Decomposition theorem of matrices
(SVD)\cite{recipes}, which states  that  given (here $m<n$, $m=31$ and $n=\infty$)
 any  $m\times n$ matrix $M_{m\times n}$, there exist essentially unique 
 matrices $U_{m\times n}$ ($m\times n$), $W_{n\times n}$ ($n\times n$), and $V_{n\times n}$ ($n\times n$)
 such that  
\beq    
M_{m\times n}= U_{m\times n} \, W_{n\times n}\,  V^T_{n\times n}.
\eeq 
 These matrices have further properties: $W_{n\times n}$ is diagonal,  with entries bigger or equal to zero,
 $V_{n\times n}$ is orthogonal, 
\beq V_{n\times n}V^T_{n\times n}=V^T V =1_{n\times n},
\eeq
 and the nonzero columns of $U_{m\times n}$ form also an orthogonal matrix($U_{m\times n} U^T_{n\times m}=1_{m\times m}$ and 
\beq
U^T_{n\times m}U_{m\times n} = P^m_{n\times n}=\mbox{diag}(1,1,..,1,0,..,0)_{n\times n}.
\eeq
There as many nonzero columns as the rank of $M$ as there are nonzero diagonal entries in $W$.

 The power of this decomposition theorem  is that it tells us what the kernel and range of $A$ are: the kernel of $A$ is spanned by the columns or rows of $V$ which correspond to the 
 zero diagonal elements of $W$, and the range is spanned by the columns of $U$ which correspond to
 the nonzero diagonal elements of $W$.

\subsection{$MM^T$}
For instance, consider
\beq
MM^T=U \, W\,  V^T V \, W \, U^T = U_{31\times \infty} W^2_{\infty\times\infty} U^T_{\infty\times 31}
\eeq
Hence, one is tempted to write 
\beq
(MM^T)^{-1} = U_{m\times n} \, W^{-2}_{n\times n} U^T_{n\times n} ,
\eeq
which is ill defined, since $W$ has nonzero diagonal entries. Let us define, in this operation, that 1/0=0. That is, all the diagonal entries of $W$ which are 0 are left alone under inversion. Does this work?
\bea
 MM^T \left(MM^T\right)^{-1} &=&  U_{m\times n}\, W^2_{m\times n}\, U^T_{n \times m}\, U_{m\times n} \, W^{-2}_{n\times n}\, U^T_{n\times m} \nonumber\\
&=&  U_{m\times n} W^2_{n\times n}\,  \mbox{diag} \, (1,1,..,1,0,..,0)_{n\times n} \, W^{-2}_{n\times n}\, U^T_{n\times m}\nonumber\\
&=&  U_{m\times n}  \, \mbox{diag}(1,1,..,1,0,..,0)_{n\times n} \, U^T_{n\times m}\nonumber\\
&=& 1_{m\times m}
\eea
so this is indeed the inverse.

\subsection{$M^TM$}

What about $M^TM$? This is
\bea
M^TM &=& V_{n\times n} \, W_{n\times n}\, U^T_{n\times m}\,   U_{m\times n} \, W_{n\times n}\,  V^T_{n\times n} \nonumber\\
= V_{n\times n} \, W^2_{n\times n}\,   V^T_{n\times n}.
\eea
This is a large matrix of rank 31. 

\subsection{Projection operators: $P=M^T   \left(MM^T\right)^{-1} M $}
Now, 
\bea
 M^T   \left(MM^T\right)^{-1} M &=&  V_{n\times n}  \, W_{n\times n}  \, U^T_{n\times m}    \;  U_{m\times n} \, W^{-2}_{n\times n} U^T_{n\times m}\;       U_{m\times n} \, W_{n\times n}\,  V^T_{n\times n} \nonumber\\
&=& V_{n\times n}  \, W_{n\times n}  \,  W^{-2}_{n\times n} \, W_{n\times n}\,  V^T_{n\times n} \nonumber\\
&=& V_{n\times n}  \,\mbox{diag}(1,1,..,1,0,..,0)_{n\times n}\,  V^T_{n\times n} \nonumber\\
&\equiv& P_{n\times n}. 
\eea
This projection operator maps the columns of $V$ corresponding to nonzero entries in the diagonal of $W$ to themselves (acts as the identity), and to zero the other ones. It projects, therefore, the linear $n$-space into and $m$-subspace.

\subsection{Projection operators: $\bar{P}=1- M^T   \left(MM^T\right)^{-1} M $}
Now, recall,
$$
Q_{n\times n}=\left[B+M^TM -M^T (MM^T)^{-1} MB  \right].
$$
But
\bea
B-M^T (MM^T)^{-1} MB &=& [1_{n\times n}-P_{n\times n} ]B \\
 &=& \bar{P}_{n\times n} B
,
\eea
and
\beq
Q_{n\times n}=\left[M^TM  +\bar{P}_{n\times n} B \right].
\eeq
note how this matrix has full rank, and how the equations have worked out so that only the ``free'' subspace of $M^TM$ is affected by the curvature equation. The minimizing set of equations are used to complement the constraints to fix a unique, well defined solution.

Let us emphasize  that $P$ and $\bar{P}= 1-P$ are both projection operators: $P^2=P$. 
\section{Complementarity: the physical and gauge sub-space}
We can now rewrite equation~\ref{eq:main}  as 
\beq
Q\cdot a = b'
\eeq
where
\beq
Q=M^T M + \bar{P}B,
\eeq
and 
\beq
b'=M^T b.
\eeq
To better interpret these terms, let us work now in the $U$ and $V$ basis. In this basis, 
\beq
Q = \left(
        \begin{array}{ccccccc}
                      w_1^2 & 0     & 0    & 0    & 0  &  ... & 0 \\
                        0   & w_2^2 & 0    & 0    & 0  &  ... & 0 \\
                        0   &   0   & ...  & 0    & 0  &  ... & 0 \\
                        0   &   0   & w_m^2& 0    & 0  &  ... & 0 \\
                        0   &   0   &   0  & 0    & 0  &  ... & 0 \\
                        ... &  ...  &  ... & ...  & ...&  ... &...\\ 
                        0   &   0   &   0  & 0    & 0  &   0  & 0 \\  
        \end{array}
    \right)
+
 \left(
        \begin{array}{ccccccc}
                        0   & 0     & 0    & 0    & 0  &  ... & 0 \\
                        0   & 0     & 0    & 0    & 0  &  ... & 0 \\
                        0   &   0   & ...  & 0    & 0  &  ... & 0 \\
                        0   &   0   & 0    & 0    & 0  &  ... & 0 \\
                        B   &   B   &   B  & B    & B  &  ... & B \\
                        ... &  ...  &  ... & ...  & ...&  ... &...\\ 
                        B   &   B   &   B  & B    & B  &   B  & B \\  
        \end{array}
    \right),
\eeq
as one would expect. In the second term we schematically show that the projection operator ``deletes'' the impact of $B$ in the physical subspace. What about $b'$? This is similarly given by
\beq
 \left(
        \begin{array}{c}
                        b'_1   \\
                        b'_2 \\
                         ... \\
                        b'_m \\
                          0 \\
                         ... \\
                          0 \\ 
        \end{array}
    \right).
\eeq 
\medskip

By ``physical'' subspace we mean the sub-vector space affected by the hard constraints associated, in this example, to the measurements we choose to respect. The complement of this sub-space is called, in other contexts, gauge sub-space.
It is the free-floating part of the solution space, the one that we need to somehow fix. \medskip

Thus, we see that the method of minimization subject to constraints leads to a nice interpretation in which the physical degrees of freedom (associated to the constraints) are not affected by the minimization extra requirement, while the gauge degrees of freedom are fixed by the curvature minimization requirement. In the language of fiber bundles, which is appropriate for handling gauge theories and which we could have used in the above discussion, by the choice of $B$ we are choosing a particular (gauge) fiber bundle section. \medskip

This method implies that we trust fully the constraints, that is, we trust fully the measurements, as if there was no noisy. This is, in general, not a good strategy, as the data may be very noisy and inconsistent. For this purpose, following our intuition  as based on the previous discussion, we can generalize a bit $Q$ to write
\beq
Q=M^T M + \bar{P}B + \nu P C,
\eeq
where $\nu$ is a ``tuning'' parameter and $C$ is a new (normalized) constraint affecting now only the physical subspace (note that tuning the middle term has zero impact). An immediate choice is $C=B$, of course.   \medskip

The first term in $Q$ is acting only in $P$-space. This encodes the limited information available from the data. The second one encodes the information from curvature minimization requirement as projected to $\bar{P}$-space. The third, new, term, encodes any additional information we may want to add to ``smooth'' or regularize the information in the first term, which may be noise or otherwise unreliable. \medskip

How much should be added to the physical part? As much as needed to fix the solution, but no more. That is, if we have some a priori knowledge on the noise characteristics associated to the $P$-space constraints, we can evaluate the entropy associated to this condition. We then need to add enough information to bring the entropy down to the desired (or needed) value.

\section{The algorithm} 
The goal is to implement code to solve
\beq
 a=\left[B+M^TM -M^T (MM^T)^{-1} MB  \right]^{-1}M^T b \equiv Q_{n\times n}^{-1}M^T b,  
\eeq
with 
\beq
Q_{n\times n}=\left[B+M^TM -M^T (MM^T)^{-1} MB  \right].
\eeq
Once the coefficients are available, a new $M$ is constructed from a interpolated set of electrode positions, $iM$. This matrix is specified here by a set of 40$\times$20 positions (the larger number for longitude positions). 

Using the interpolating matrix, it is then simple to compute the interpolated potential field and its Laplacian.

\section{Analysis of sample experimental data}
 Here we give a simple example using an  implementation of this algorithm developed at Starlab.  A  MMN sample data set from an Experiment carried out at the University of Barcelona Neurodynamics Laboratory is analyzed for illustration. For the purposes of the present discussion it suffices to mention that the 32 electrode (plus a reference) data set corresponds to an average of MMN  EEG from 17 control group subjects, and that it corresponds to 100 ms prior to an auditive stimulus up to 500 ms later.   For more information see \cite{ruffini2000} and forthcoming publications. Here we show results for lmax=20.

%\clearpage
\subsection{Control group}
This is the output for the data interpolation (32 electrodes, time in ms):\medskip

\noindent \epsfxsize=13cm \epsffile{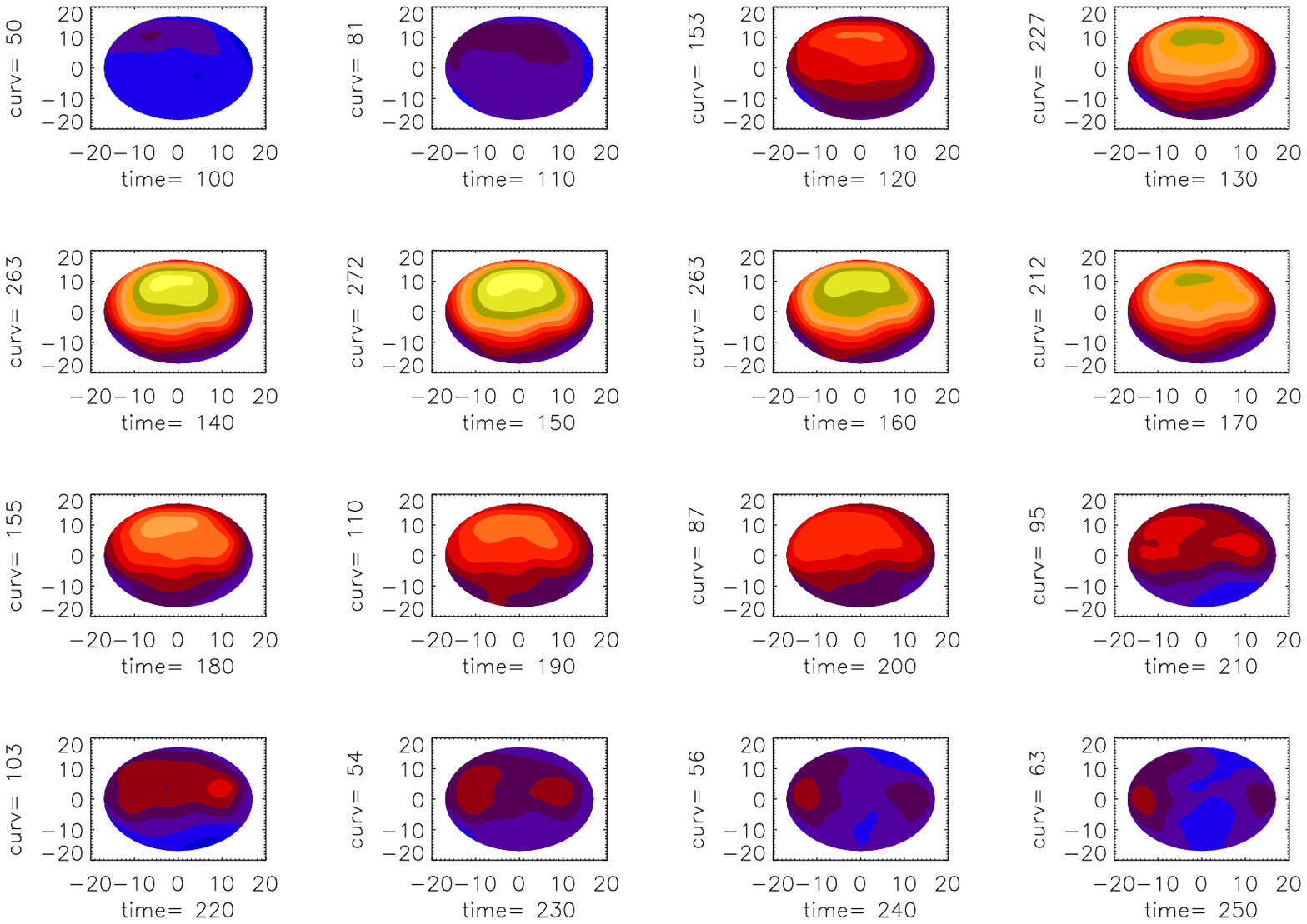} \\ \vspace{0in}

And this is the resulting Laplacian:

\noindent \epsfxsize=13cm \epsffile{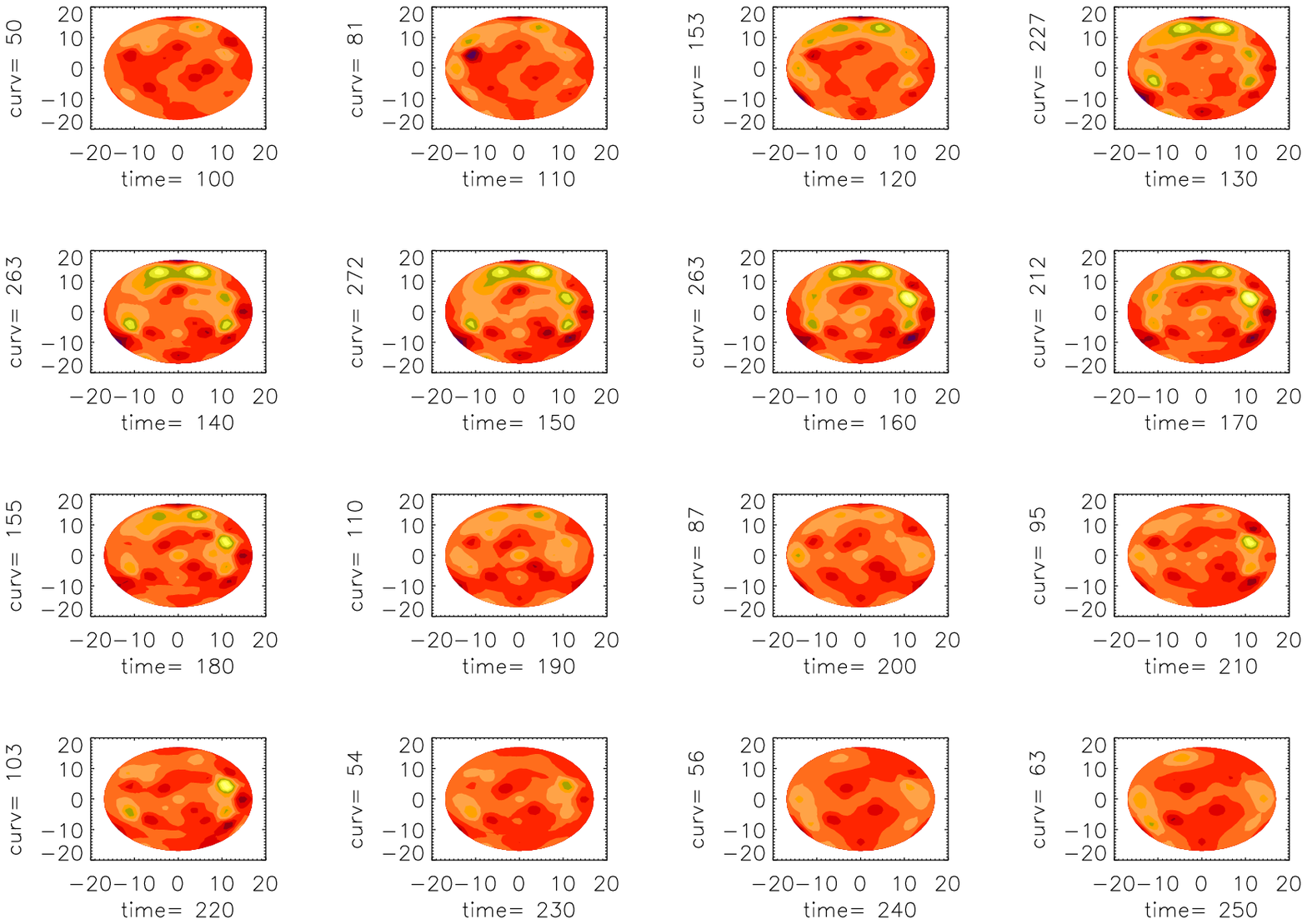} \\ \vspace{0in}

\subsection{Curvature index}
This a graph of the curvature of the fitted minimal surface, an interesting measure (the blue curve, the other curve refers to another experimental group).  

\epsfxsize=10cm \epsffile{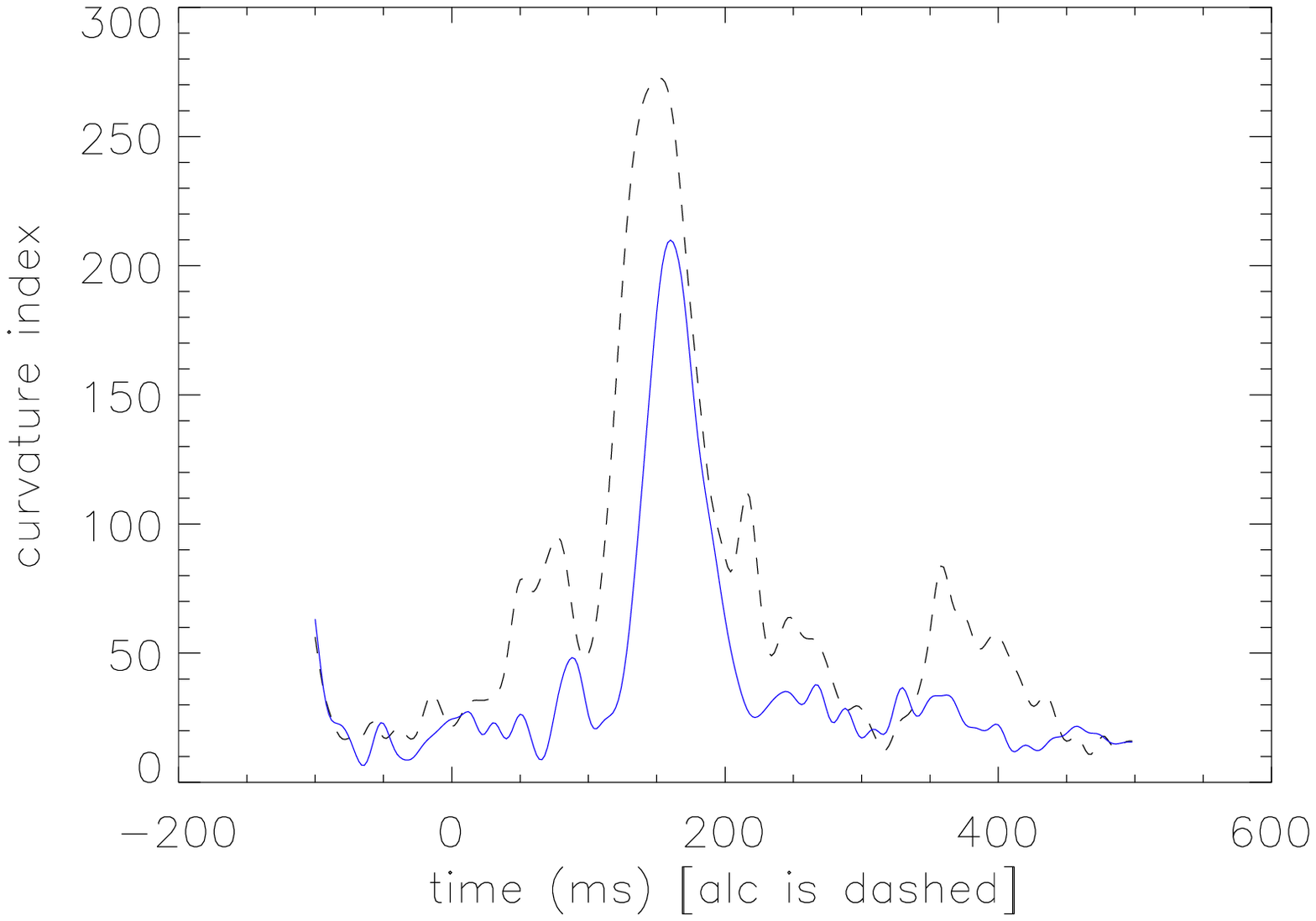} \\ \vspace{0in}

%%%%%%%%%%%%%%%%%%%%%%%%%%%%%%%%%%%%%%%%%%%%%%%%%%%%%%%%%%%%%%%%%%%%%%%%%

\section{Analogy to Constrained Mechanics and Gauge Theories}

In this set of notes I intend to introduce very quickly to the Lagrangian
formalism in physics, the impact of symmetries of the action to the solution
space, their relation to gauge theory, and how all these strange things happen
to relate to the interpolation problem (for more on this and further references, see xxx-gr-qc9806058). What all these have in
common is their origin in a minimization (or extremization, to be precise) problem.

\subsection{The Lagrangian in physics and \protect\( \chi ^{2}\protect \)}
It is interesting to form an analogy to the theory constrained theories in classical mechanics. \medskip

Given a system with \( n \)degrees of freedom, each specified by a coordinate
\( q_{i}(t) \), we can obtain its dynamics by minimizing the functional (called
the action)
\begin{equation}
S[q_{i}(t)]=\int _{t_{1}}^{t_{2}}L(q_{i}(t),\dot{q}_{i}(t))\: dt,
\end{equation}
 that is, one looks for the set of \( q_{i}(t) \)such that given that at \( t_{1} \)
and \( t_{2} \)the values of the coordinates are to be held fixed, one varies
the functions so as to find the minimum of \( S \). In general, and although
this may seem less familiar to you, one can rewrite this as
\begin{equation}
S[q_{i}(t)]=\int _{t_{1}}^{t_{2}}\int _{t_{1}}^{t_{2}}q_{i}(t)M^{ij}(t,t')q_{j}(t')\: dtdt'-\int _{t_{1}}^{t_{2}}V(q_{i}(t))\: dt
\end{equation}
 If the last term is a quadratic (as it happens in the case of doable physics,
e.g., the harmonic oscillator), we find that we are trying to minimize a quadratic
functional, just as was the case in the previous sections, where the goal was to minimize the curvature. There is no single surface of minimal curvature, as a constant term can always be added, for example. \medskip

Allowing for linear terms, then, we are trying to minimize something
that looks like
\begin{equation}
S=q^{T}Mq+q\cdot j+c
\end{equation}
where \( q \) here stands for a long vector in which there is an entry for
each time parameter (think of time as a discrete index if this helps)and for
each possible value of the index \( i \). By a symmetry we mean a transformation
of the quantities \( q\rightarrow q' \)so that \( S\rightarrow S \): i.e.,
so that the action is invariant. For instance, and to simplify things, imagine
that the action we want to minimize is 
\begin{equation}
\chi ^{2}(x)=(y-Ax)^{2}
\end{equation}
where \( x \), which we will call here the \emph{state vector} or \emph{state},
for short, is an \( n \)-vector, and \( y \) is an \( m \)-vector (this equation
may look familiar to you!). Let us now use the Singular Value Decomposition
theorem to rewrite 
\begin{equation}
A=UWV^{T}
\end{equation}
Recall that \( W \) is a \( n\times n \) diagonal matrix with entries bigger
or equal to zero, \( V \) is an orthogonal \( n\times n \)-matrix (\( VV=^{T}VV^{T}=I_{n\times n}), \)
and that \( U \) is and \( m\times n \)-matrix with orthogonal columns (\( U^{T}U=I_{n\times n}) \). \medskip

 Imagine now that the last \( m \) diagonal
entries of \( W \) are zero. What does this mean? It means that the two states
\( x \) and \( x+\alpha _{j}v^{j} \)are equivalent as far as \( \chi ^{2} \)
is concerned, where by \( v_{j} \) we mean the \( j \)-th column of \( V \):
\begin{equation}
\chi ^{2}(x)=\chi ^{2}(x+\alpha _{i}v^{i}),\: \; \: i=1,...,m
\end{equation}
 The fact that the \( m \) coefficients \( \alpha _{i} \) are all independent
of each other is what makes this transformation symmetry a \emph{local symmetry}
in the physics lingo. It tells us that of the original \( n \) degrees of freedom,
only \( n-m \) really matter. These are called the \emph{physical degrees of
freedom}. The rest are called \emph{gauge} or \emph{unphysical}, degrees of
freedom. These are the source of our inversion problems: the minimization problem
does not have a unique solution. In physics one fixes this in some way or another,
usually by adding a so-called \emph{gauge-fixing} conditions. This means to pick
a set of coefficients \( \alpha _{i}, \) and this is usually done by adding
a set of \( m \) equations to the problem that fix these. Let \( B \) be a
rank-\( m \) \( n\times n \)-matrix. Then we'd like to add such a condition
to the problem. Notice the condition \( Bx=0 \) is to fix the gauge degrees
of freedom then \( B \) has to have maximal rank in the gauge subspace (i.e.,
it must completely span the gauge subspace).  \medskip

 If we believe in the data, one must avoid increasing \( \chi ^{2} \) while fixing the gauge degrees of freedom.
It is possible, as we now discuss, to fix these in a reasonable manner,
without getting away from the data. We have in fact discovered how to do that in the previous discussion, but it may be useful to repeat the exercise in this context. \medskip

Suppose then that for some physical reason we want to ask that \( Cx=0 \) in
order to fix the gauge degrees of freedom, but that we have been a little naive
and have not worried about the rank of \( C \). Proceeding as is usual,
then, we choose to minimize the functional 
\begin{equation}
\chi ^{2}(x)=(y-Ax)^{2}+\lambda (Cx)^{2}.
\end{equation}
Upon minimization we obtain the equation
\begin{equation}
(A^{T}A+\lambda C^{T}C)x=A^{T}y.
\end{equation}
Let \( x=Vx'. \) This is the state in ``SVD coordinates''. The first \( n-m \)
entries in the vector \( x' \) are physical coordinates, the rest are gauge.
Then, using the fact that \( A^{T}A=VW^{2}V^{T} \) , we can rewrite this equation
as 
\begin{equation}
V^{T}(A^{T}A+\lambda C^{T}C)VV^{T}x=V^{T}A^{T}y,
\end{equation}
or

\begin{equation}
(W^{2}+\lambda V^{T}C^{T}CV)x'=WU^{T}y.
\end{equation}
This equation reflects the fact that without the additional constraint (i.e.,
set \( \lambda =0 \)), the physical degrees of freedom are well-fixed, while
the gauge ones are not. If we want to avoid disturbing the physical part of
the state, we now project the constraint into the gauge subspace: \( V^{T}C^{T}CV\rightarrow P_{gauge}V^{T}C^{T}CV \),
where \( P_{gauge} \) is a diagonal matrix with zeros everywhere except the
last \( m \) last diagonal elements. Now the solution to
\begin{equation}
(W^{2}+\lambda P_{gauge}V^{T}C^{T}CV)x'=WU^{T}y
\end{equation}
is as before as far as the physical degrees of freedom are concerned. If the
rank of \( P_{gauge}V^{T}C^{T}CV \) is \( m \) we will have fixed a complete
solution. To see more clearly what has happened, let us factorize \( x'=x_{phy}+x_{gauge} \),
where \( x_{gauge}\sim \alpha _{i}v^{i} \) is the projection of \( x \) into
the gauge subspace , we can rewrite the above equations as 
\begin{equation}
W^{2}x_{phys}=WU^{T}y,
\end{equation}
 which fixes the physical degrees of freedom , as it should, and

\begin{equation} 
P_{gauge}V^{T}C^{T}CV\left( x_{phy}+x_{gauge}\right) =0.
\end{equation}
In this way we ensure that we are not imposing any constraint equations beyond
those that we are really allowed to. The matrix \( P_{gauge}V^{T}C^{T}CV \)
is to have rank \( m. \) If \( C \) is fully ranked, then it will. If not,
we may still be ok. We can now go back to the original coordinates and write
(notice that \( \lambda  \) is irrelevant now)
\begin{equation}
(A^{T}A+VP_{gauge}V^{T}C^{T}C)x=A^{T}y
\end{equation}
Another way to reason this result is the following. As we mentioned, the physical
degrees of freedom, \( P_{phys}x \), are fixed. We want to really find now
the set of gauge degrees of freedom such that (\( C'=V^{T}CV \) is the expression
of the constraint in the SVD coordinates)
\begin{equation}
\left( C'(P_{gauge}x'+P_{phys}x')\right) ^{2}
\end{equation}
is minimum. If we vary the gauge degrees of freedom in this functional and set
it equal to zero we find the equation
\begin{equation}
\left( P_{gauge}C'^{T}C'P_{gauge}+P_{gauge}C'^{T}C'P_{phys}\right) x'=0,
\end{equation}
or \( P_{gauge}C'^{T}C'x'=0 \), since \( P_{gauge}+P_{phys}=I_{n\times n} \).
In the original coordinates, this reads \( VP_{gauge}V^{T}C^{T}Cx=0 \), as
we wrote before. 

By construction, then, there exists now a unique solution to the problem
\begin{equation}
\left( A^{T}A+VP_{gauge}V^{T}C^{T}C\right) x=A^{T}y.
\end{equation}
 %This is the set of equations we should really solve. In an algorithmic sense,
%the procedure is as follows. You start with a matrix 
%\begin{equation}
%M_{n}=A_{n}^{T}A_{n}+\left( C_{n-1}+\delta ^{2}\right) ^{-1}
%\end{equation}
%which encodes the data from a batch and the covariance information from the
%previous batch. One must then compute \( M_{n}=U_{n}W_{n}V_{n}^{T} \), determine
%the gauge projector \( P_{gauge} \) (really just count the nonzero diagonal
%entries in \( W \), since \( P_{gauge} \)is a diagonal matrix with unit entries
%and rank \( m \)), and then solve
%\begin{equation}
%\left( M_{n}+V_{n}P_{gauge,n}V_{n}^{T}C^{T}C\right) x=A_{n}^{T}y_{n}+\left( C_{n-1}+\delta ^{2}\right) ^{-1}x_{n}.
%\end{equation}
%In terms of what we have already implemented, the change is simple: just change
%\( C^{T}C \) for \( V_{n}P_{gauge}V_{n}^{T}C^{T}C \). And forget about \( \lambda  \),
%it cannot have any impact on the solution anyhow.

%What about fiber bundles? Well, the solution space of the ill-posed tomographic
%problem \emph{can} a fiber bundle: a trivial fiber bundle, where the fiber space
%is spanned by the gauge degrees of freedom, and the structure group is rather
%stupid : \( Z_{1}={1}. \)


\begin{thebibliography}{1}

\bibitem{recipes}
S~A~Teukolsky, W~H~Press, W T~Vettering,  Flannery,
\newblock {\em  Numerical Recipes in Fortran, The Art of Scientific Computing},
\newblock  Cambridge University Press, 1994.


\bibitem{ruffini2000}
 Ruffini, G., Galan, F., Marco, J., Polo, MD., Escera, C., Rius, A., Grau, C., Estudio de patrones temporo-espaciales de activacion cerebral durante la produccion de Mismatch Negativity en el alcoholismo cronico, Santiago de Compostela, 21st September, 2000. Abstract available at http://www.starlab.es.

\bibitem{ruffinithesis} Ruffini, G., PhD thesis, The Quantization of Simple Parametrized Theories, UC Davis, 1995 (available at http://www.starlab.es).

\end{thebibliography}
\end{document}